\documentclass[12pt]{article}

\usepackage{arxiv}

\usepackage[T2A, T1]{fontenc}
\usepackage[utf8]{inputenc}
\usepackage[russian, english]{babel}

\usepackage{textcomp}
\usepackage{gensymb}
\usepackage{authblk}
\usepackage{xcolor}
\usepackage{graphicx}
\usepackage{amsmath}
\usepackage{hyperref}

\usepackage{csquotes}
\usepackage[
    backend=biber,
    style=alphabetic,
    sorting=ynt,
]{biblatex}
\graphicspath{ {./images/} }

\providecommand{\keywords}[1]{
  \small
  \textbf{\textit{Keywords---}} #1
}

\makeatletter
\renewcommand\AB@affilsepx{, \protect\Affilfont}
\makeatother

\title{Raccoons vs Demons: \\multiclass labeled P300 dataset}
\author[1, 3]{Goncharenko V.}
\author[1, 2]{Grigoryan R.}
\author[1, 3]{Samokhina A.}
\affil[1]{Neiry}
\affil[2]{MSU}
\affil[3]{MIPT}
\date{\today{}}

\begin{document}

\maketitle

\begin{abstract}

    We publish dataset of visual P300 BCI performed in Virtual Reality (VR) game Raccoons versus Demons (RvD). Data contains reach labels incorporating information about stimulus chosen enabling us to estimate model's confidence at each stimulus prediction stage.
    % Also we consider our dataset in comparison to existing open datasets and apply transfer learning.

    Data and experiments code are available at \url{https://gitlab.com/impulse-neiry_public/raccoons-vs-demons}

\end{abstract}


\section{Introduction}

    Firstly Brain-Computer Interfaces (BCI) were developed mostly with disabled patients in mind (e.g. \cite{BCI_Competition_III}, \cite{bnci-horizon-2020-8}, \cite{EPFL_P300_dataset}). Nowadays there are numerous attempts to employ BCI solutions for healthy people using noninvasive electrodes (see \cite{NER_2015}, ).

    More than that BCI numerously applied to recreational activities such as games \cite{Recreational_Applications}, \cite{Kaplan_Shishkin_games}

    Another crucial thing for further BCI development is processing and visualization instruments. At the beginning every laboratory used own programming environments and data format, some even used proprietary platforms. Happily at the moment we have fully open sourced systems based on Python programming language like MNE \cite{MNE_package}, PyRiemann \cite{PyRiemann} and MOABB \cite{MOABB_package}, also EDF data format \cite{EDF_format} to store EEG/MEG data specifically.

\section{Materials and methods}

    \subsection{Participants}

        61 healthy participants (23 males) naive to BCI with mean age 28 years from 19 to 45 y.o. took part in the study.

        All subject signed informed consent (see appendix) and passed primary prerequisites on their health and condition.

    \subsection{Stimulation and EEG recording}

\begin{figure}
            \centering
            \includegraphics[width=200pt]{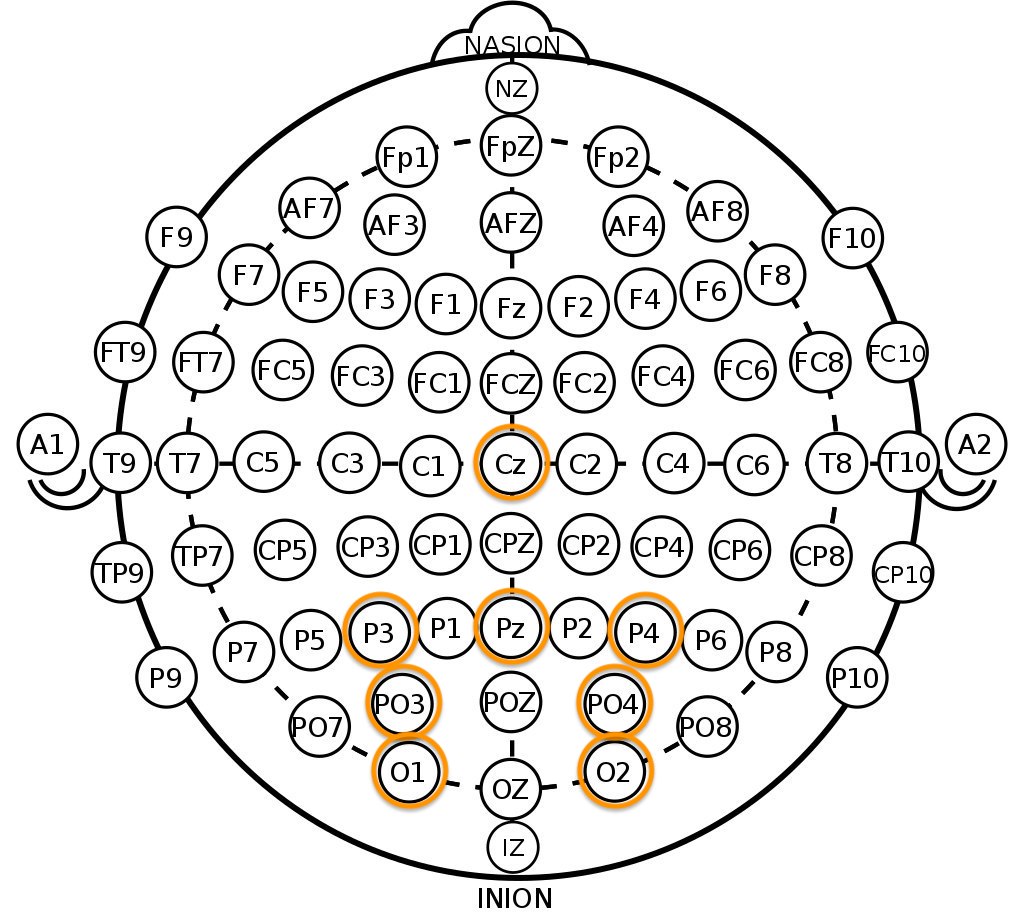}
            \caption{Electrodes poisitions}
            \label{electrodes}
        \end{figure}

        The EEG was recorded with NVX-52 encephalograph (MCS, Zelenograd, Russia) at 500 Hz. We used 8 sponge electrodes (Cz, P3, P4, PO3, POz, PO4, O1, O2), see \ref{electrodes}. Stimuli were presented with HTC Vive Pro VR headset with TTL hardware sync.

    \subsection{Experimental procedure}

        Participants were asked to play the P300 BCI game in virtual reality. BCI was embedded into a game plot with the player posing as a forest warden. The player was supposed to feed animals and protect them from demons.

\begin{figure}
            \centering
            \includegraphics[width=300pt]{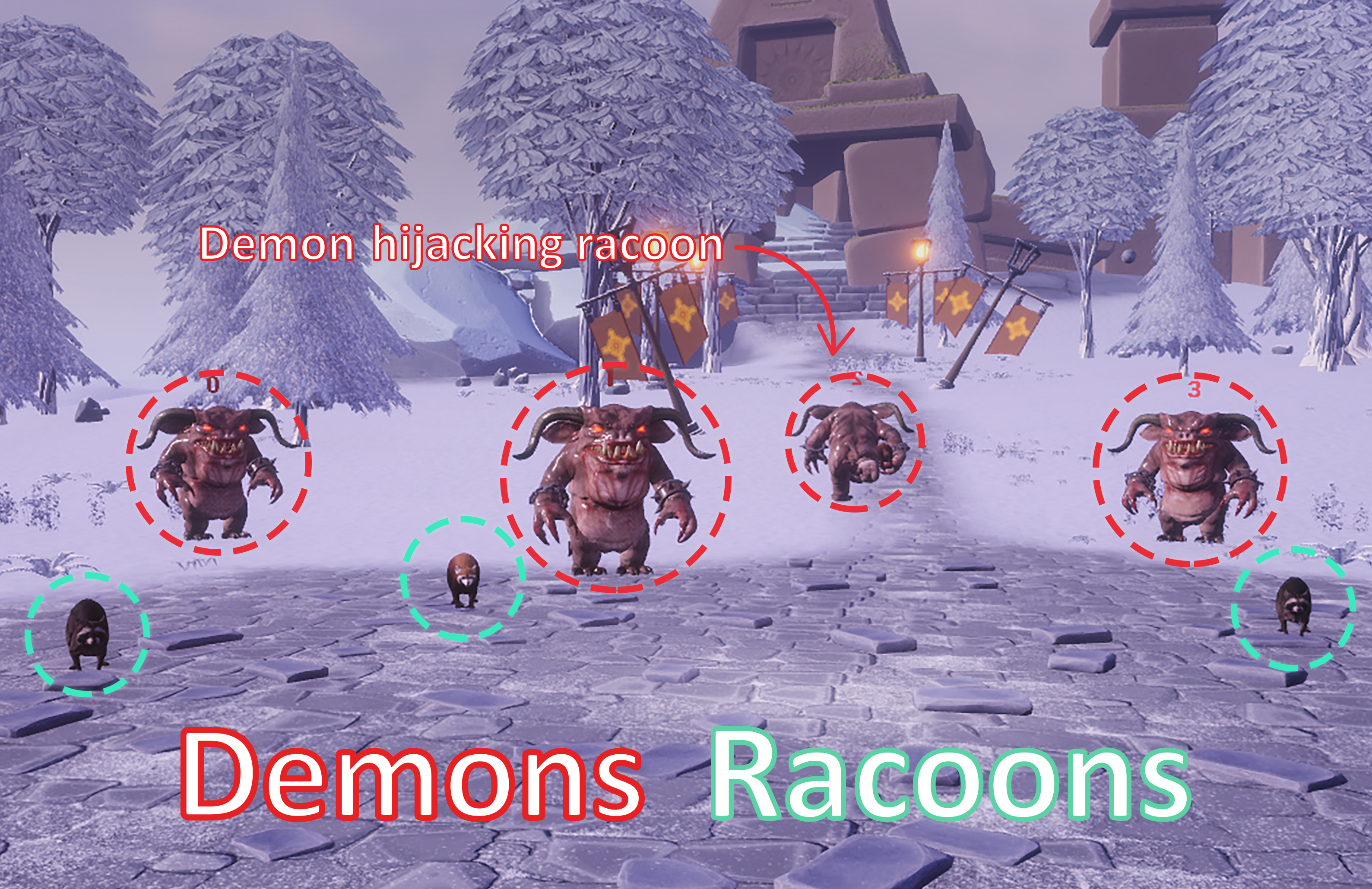}
            \caption{Scene from VR game showing raccoons and demons}
            \label{gameplay}
        \end{figure}

        We have used an oddball paradigm. At the learning stage, 5 numbered animated raccoons (see \ref{gameplay}) were presented to the participant. The stimulation consisted of raccoon jumps in a random order to prevent activation anticipation (for animation timeline see fig. 2B). Participants were instructed to count jumps of a specific raccoon for 5 sessions. Each session consisted of all targets activating 10 times, after which the animal received food. This produced 50 target and 450 non-target evoked potentials that were used for classification.

        During BCI feedback scene, stimuli were presented as animated demons, which were jumping. The instruction for each participant was essentially the same as for the learning session (counting jumps of the target stimuli). Before each session the participants called out the number of demon they would be counting during the session. This was done to get rid of the usual copy-spell scheme, and to preserve interactivity (participant choose target himself). Selected demon was destroyed at the end of the scene (10 activations of all stimuli). Then demons came closer to the participant, eventually grabbing one of the raccoons and running away. When the target was removed from the screen one way or another, the new demon was spawned to the initial position, so the number of stimuli remained constant.

        Data was recorded for later offline analysis.
        The game was accompanied with spoken instruction and background music.

\section{Data structure and terms}

    Essentially in an oddball paradigm we have to solve 2 different but connected classification tasks:
    \begin{enumerate}
        \item Binary

        Classification of a single epoch to contain P300 or not (in other words to be target or empty).
        In this case each epoch is classified independently. It is suitable for online (calibration then test on the same person) training because only 250 epochs are enough to train such a model. Another noticeable trait of this type of problem is imbalanced training and testing sets because each stimulus is activated the same number of times while only one of them is target. So we end up with $1$ to $(s-1)$ class ratio where $s$ is the number of stimuli.
        For this task accuracy metric is not sufficient because of class imbalance so we track precision, recall, f1 and ROC AUC in all experiments.

        \item Multiclass

        Classification of the stimulus chosen by user. As long as we have several stimuli (usually 5 to 7) it's a multiclass problem. Here we consider several epochs at a time (usually equal number from each stimulus) and have prior knowledge that only one stimulus could be the target thus others are empty. In this case classification is more balanced because the participant chooses different stimuli each time (in fact it depends on visualization setup). Usually this problem is solved by summation of probabilities of binary problem by stimulus and choosing the one with highest sum.
        Regarding metrics this task is well described by accuracy.

    \end{enumerate}

    \begin{figure}
        \centering
        \includegraphics[width=300pt]{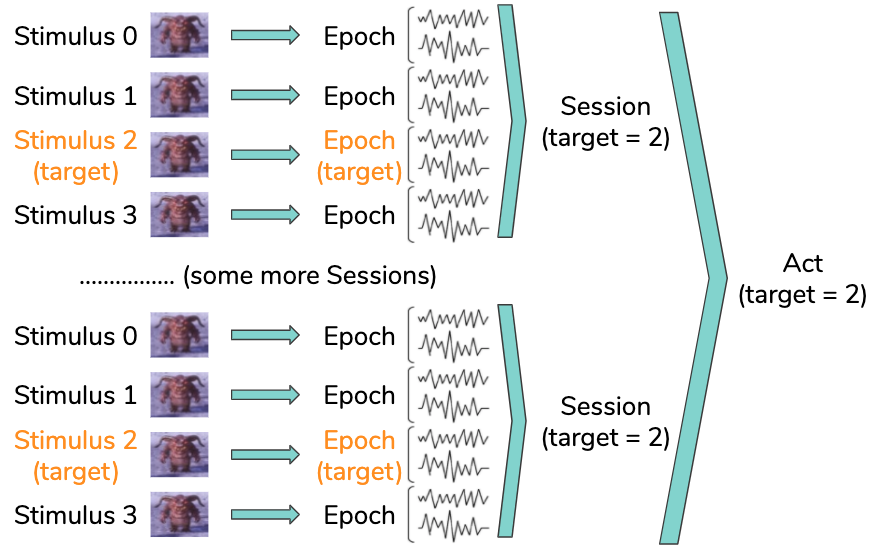}
        \caption{Data hierarchy scheme}
        \label{data_structure}
    \end{figure}

    Let's introduce some terminology helpful in further narration (for illustration see \ref{data_structure}:

    \begin{itemize}
        \item Round - sequence of epochs with one target where each stimulus was activated once (and produced one epoch), so this is quantum of multiclass prediction. Usually stimuli are activated in random order.

        \item Act - sequence of rounds with one target after which decision on target stimulus is made. Usually the number of sessions is from 5 to 10 in our experiments.

        \item (Game) Record - sequence of acts obtained from one person during one recording session (one game).
        
        \item Dataset - collection of records sharing the same visual activation and hardware setup.
    \end{itemize}

\section{Preprocessing and Classification}
    
    As a main metrics we use multiclass accuracy because in a final gaming environment this is how user measures performance. Binary task metrics are measured in all experiments as well and in many cases higher binary metrics means higher multiclass accuracy.

    We divide EEG signal handling to two principal steps:
    \begin{enumerate}
        \item Preprocessing - which includes signal processing stuff and some non-statistical permutations e.g. clipping, epochs slicing, etc. At the end of this step we obtain epochs with labels (either binary or multiclass).
        \item Classification - applying statistical classifiers to predict labels having epoch data
    \end{enumerate}
    
    % Classification methodology

    For preprocessing step we search for optimal hyperparameter values by cross-validation. The reason for this seemingly trivial step is lack of motivation in choosing such parameters. Although many researches report similar filtering (e.g. \cite{Brain_Invaders_2011}, \cite{Plug_and_Play_P300_BCI}, \cite{How_many_people_are_able}, \cite{BCI_competition_III_svm}, \cite{MI_dataset}) they don't describe motivation for choosing one or another downsampling factor, filtering band, clipping as well as epoch duration. Below we report results of classification pipeline with varying parameters and choose ones that minimize final epoch representation without metrics degradation.
    
    For this step we use predefined set

    Binary epochs classification is a common task for BCI and main approaches are well described in \cite{Lotte_2018}. In recent years Deep Learning emerged in all fields, so EEG analysis is not an exception and we have beautiful overviews of \cite{Craik_2019} and \cite{Roy_2019}.

    Our preprocessing is quite standard:
    \begin{enumerate}
        \item Decimation with an anti-aliasing filter from SciPy package \cite{SciPy}
        \item Bandpass filtering with Butterworth IIR
        \item Clipping off scale samples
        \item Channelwise standard scaling (subtracting mean and dividing by std for all values of each channel)
    \end{enumerate}

    \begin{figure}
        \centering
        \includegraphics[width=300pt]{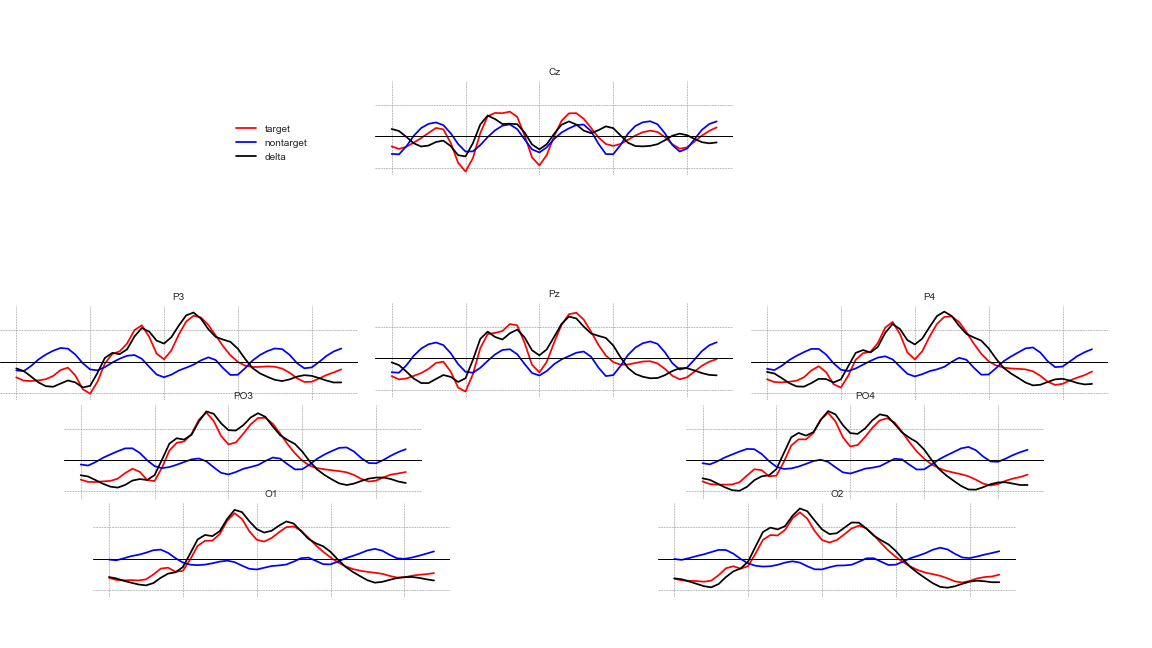}
        \caption{Average P300 (target) and non-P300 (empty) responses by electrodes}
        \label{p300_responses}
    \end{figure}

    % Usually researchers use epoch duration 300-1000 ms to predict P300 response. We have decided to obtain optimal interval in order to reduce initial data dimensionality and potentially reduce time needed to perform one stimulus prediction. More than that epoch 
    
    Each test were conducted with other parameters fixed at some sane level (such that qualitative result persisted)
    
    \subsection{Decimation}

\begin{figure}
            \centering
            \includegraphics[width=300pt]{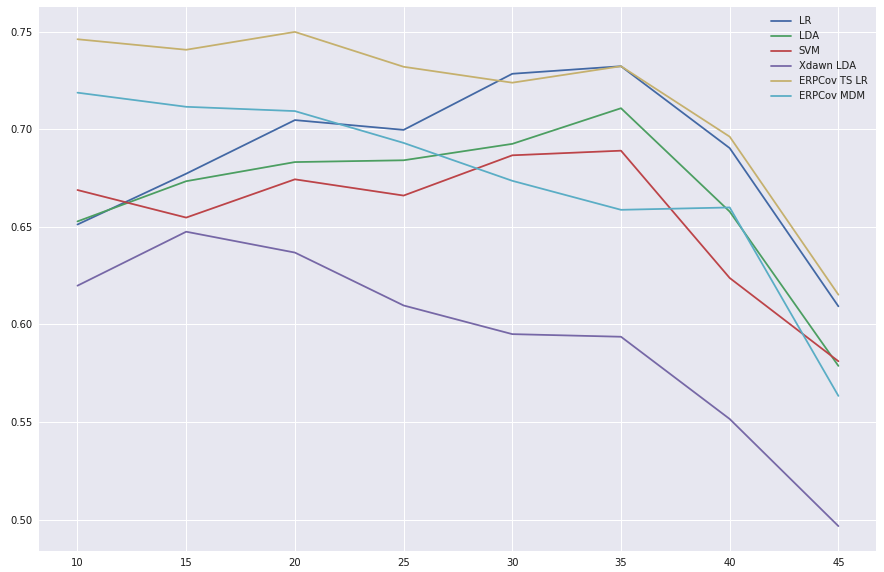}
            \caption{Dependency of final prediction accuracy on decimation factor}
            \label{decimation_accuracy}
        \end{figure}

        First we need to determine which minimal signal frequency we will operate on in all later stages. On \ref{decimation_accuracy} we see that classification degrades for decimation more than 35 times resulting in a rate of about 15Hz. So this is the upper value of frequency on which P300 occurs. Different lines correspond to different final classifiers and transparent zones of corresponding colour refer standard deviation by persons.

    \subsection{Filtering}

\begin{figure}
            \centering
            \includegraphics[width=300pt]{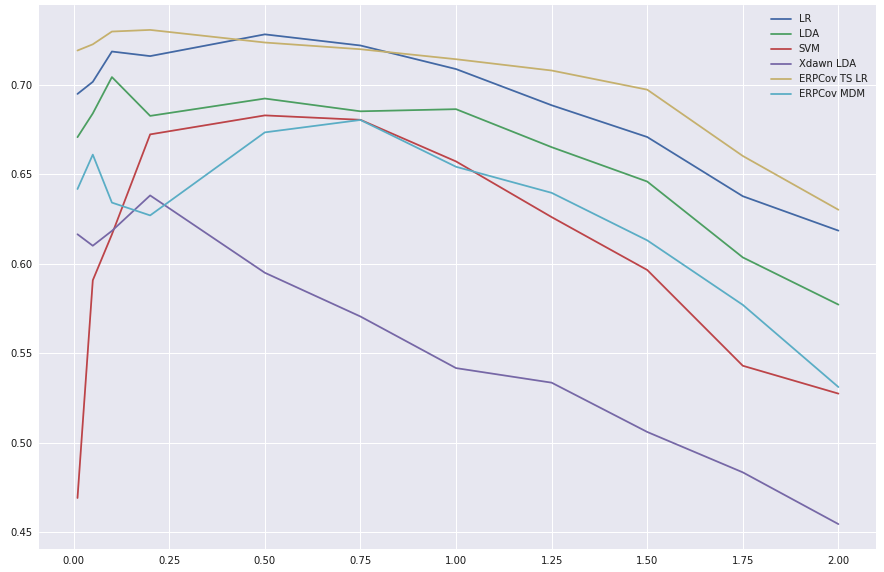}
            \caption{Dependency of final prediction accuracy on lower filtering bound}
            \label{lowfreq_accuracy}
        \end{figure}

        Since we optimized decimation and chose minimal available frequency then we only need to filter low frequencies with Butterworth filter. On \ref{lowfreq_accuracy} we see that filtering very low frequencies (below 0.2Hz) is crucial for classification and band from 0.2 to 1 doesn't affect final classification quality.

    \subsection{Epoch duration}

        The same operation was held over epoch durations. First we tested right bounds and found quality degradation on 600ms (see \ref{epoch_right_accuracy}).

        \begin{figure}
            \centering
            \includegraphics[width=300pt]{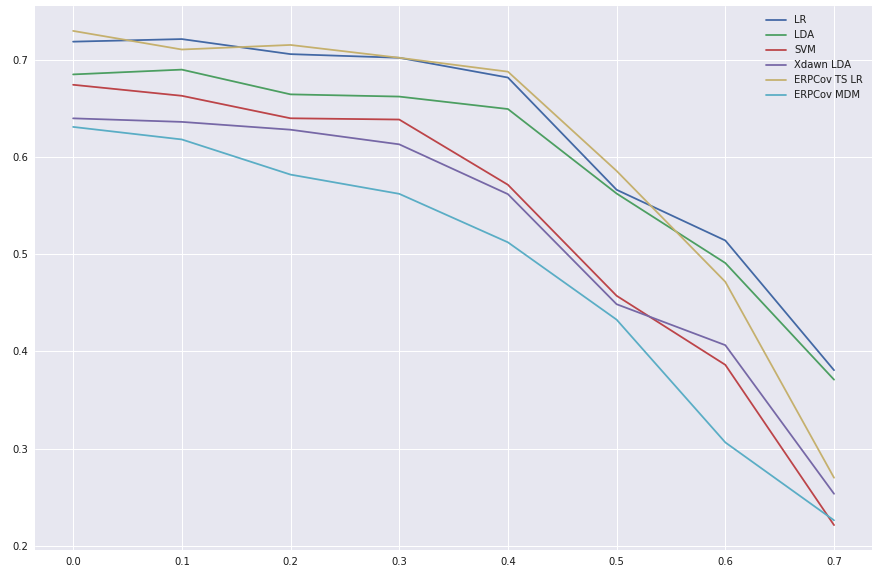}
            \caption{Dependency of final prediction accuracy on left bound of epoch}
            \label{epoch_left_accuracy}
        \end{figure}

        \begin{figure}
            \centering
            \includegraphics[width=300pt]{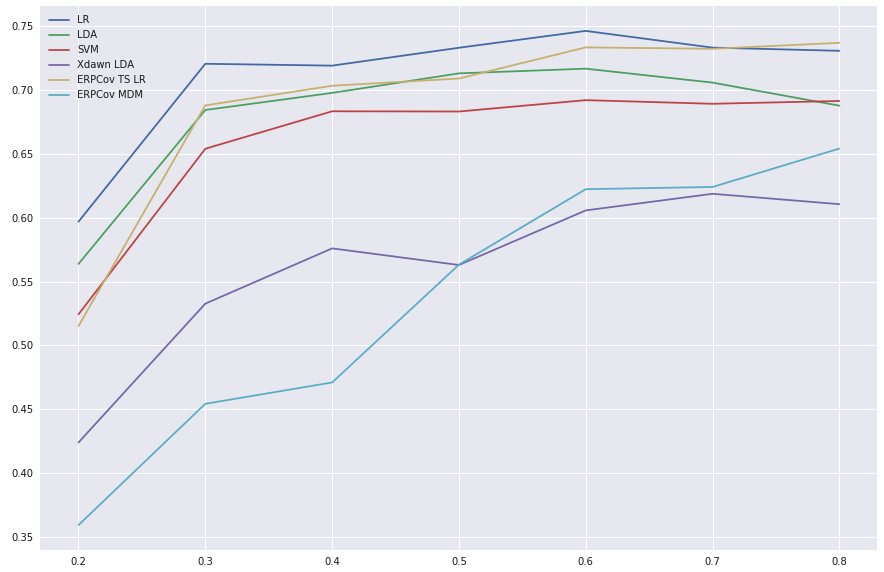}
            \caption{Dependency of final prediction accuracy on right bound of epoch}
            \label{epoch_right_accuracy}
        \end{figure}

\section{Multiclass confidence level}

    Having robust binary target-empty epoch classifier (trained for each person separately) we may proceed to aggregating these predictions to a final multiclass problem solution.
    
    As a standard approach we consider taking epochs grouped by activated stimulus, predicting their probabilities to be target and sum these probabilities. This way we obtain \emph{"activation score"} for each stimulus. Then multiclass prediction consists in choosing the stimulus with the highest activation score. We call it baseline multiclass classification.
    
    In gaming environment (which is our case) misclassification cost is high because person feels frustrated and snatched. So we have decided to introduce some confidence score by which we could reject potentially wrong predictions. In case of rejection "misfire" animation is played and we ask the user to repeat the choice procedure.
    
    Some attempts to fix classification errors were taken. For example in \cite{NER_2015} competition EEG signal itself used to predict if SVM classifier were right or not.
    
    \subsection{Heuristics}

        This score could be based on prior knowledge that one act has only one target. Having activation scores of each stimulus we expect a good binary classifier to give low scores to all but one stimulus. And distribution of sorted activation scores confirms this statement.

        \begin{figure}
            \centering
            \includegraphics[width=300pt]{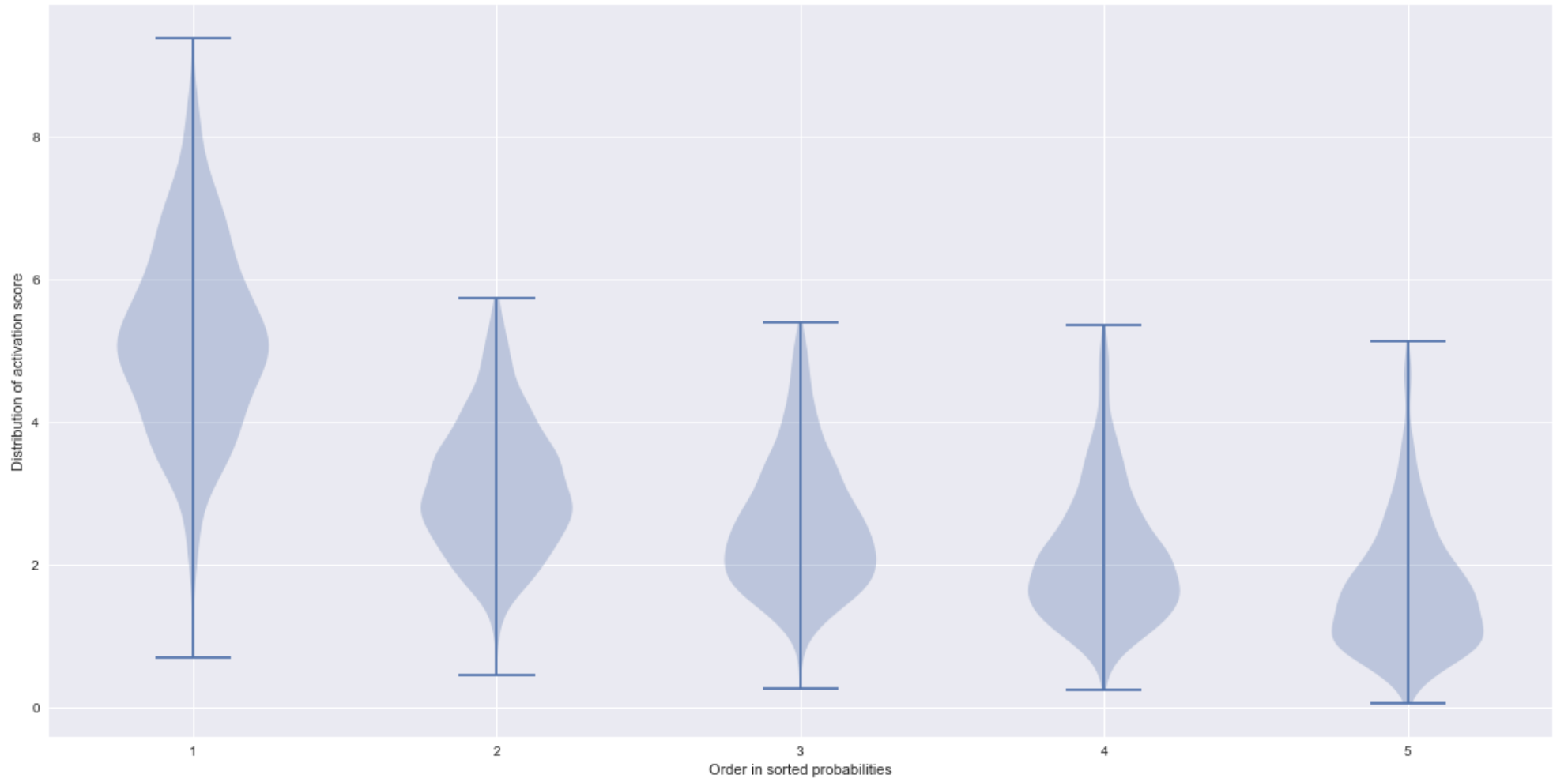}
            \caption{Distribution of sorted activation scores}
        \end{figure}

        When there isn't one obvious leader it's a motive to suspect lower probability to give the right answer. So our task is to find some criterion (which we call confidence score) that separate wrongly classified acts.
        During experiments we have considered different confidence scores based on activation scores:

        \begin{itemize}
            \item sum of activations
            \item max activation
            \item mean activation
            \item median activation
            \item min activation
            \item max - mean
            \item max - median
            \item second max
            \item max - second max
            \item max - mean of all but max
            \item max - median of all but max
            \item max - sum of activations
        \end{itemize}

        To estimate the threshold value we plot criterion distribution for correctly classified acts and for misclassified ones.

        \begin{figure}
          \centering
          \includegraphics[width=300pt]{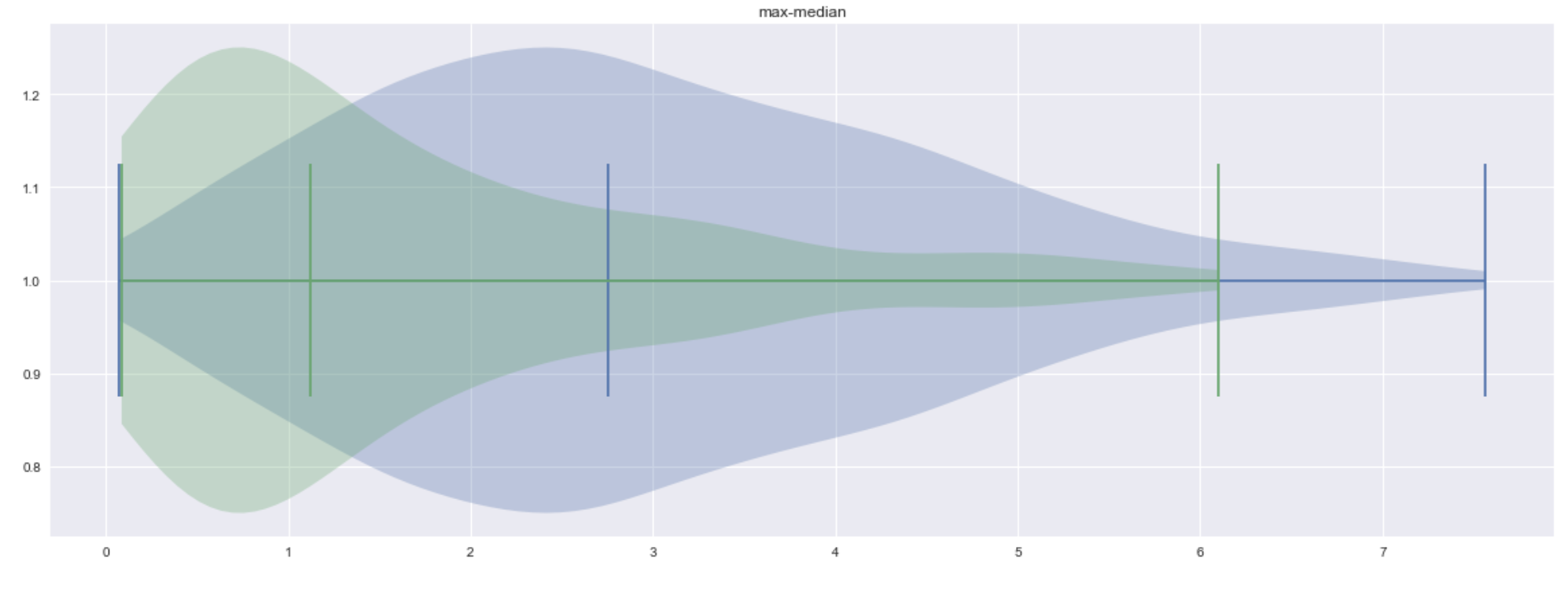}
          \caption{Confidence score distribution for 'max-median' heuristic}
        \end{figure}

        On average confidence scores are lower for misclassified acts, so we are able to pick those acts among which the rate of correctly classified (precision) is high. But considering the game setup we couldn't afford to trigger "misfire" action too frequently because it would be too disturbing for a user. We have stated the rejection rate couldn't be more than 30\%, precisely 10\% to 20\% would be optimal because the user encounters misfire at some point but doesn't get overwhelmed by this mechanics.

        Finally we can see that 'max-mean' and 'max-median' performs best at 5 - 50\% of rejected acts which is the region of interest for us (see \ref{negrate_precision_curve}). For the best heuristic 'max-median' we could increase accuracy by 5\% sacrificing only 12\% of all acts on average. However this misfiers is concentrated in a few participants which is evidenced by a median accuracy of 90\% in this case.

    \subsection{Stacking}

        In fact we perform some kind of stacking, so we may apply a model to predict if this act could be misclassified.

        Curve of negative rate vs precision in 5-fold cross-validation is shown on plot for Logistic Regression, Random Forest and our Heuristics Threshold classifiers.

        \begin{figure}
            \centering
            \includegraphics[width=300pt]{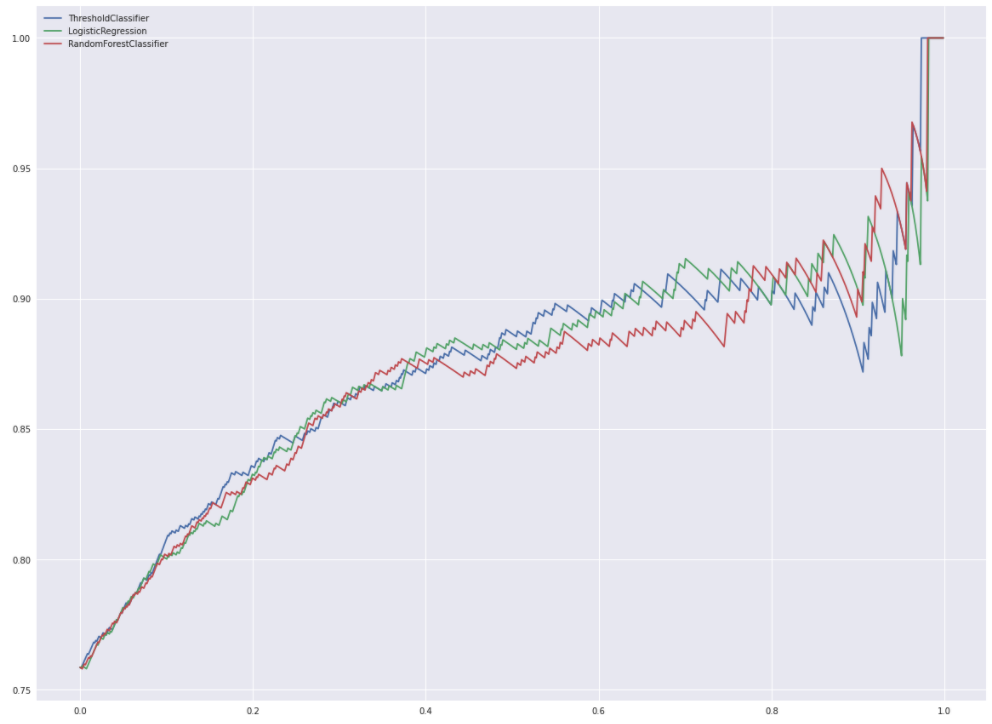}
            \caption{Negative rate vs Precision for different models}
            \label{negrate_precision_curve}
        \end{figure}

        This plot shows that in range 10 to 25 \% our heuristics work the same as trained classifiers and provides 4 to 9 \% increase in precision of predictions (see \ref{negrate_precision_curve}).

\section{Discussion}

    Introduction of multiclass labels allows us to incorporate prior knowledge of choosing only one stimulus per session. This could efficiently be used in Neural Networks architecture development as well as in Bayesian methods based approaches.

\printbibliography

@article{
    Lotte_2018,
	doi = {10.1088/1741-2552/aab2f2},
	url = {https://doi.org/10.1088%2F1741-2552%2Faab2f2},
	year = 2018,
	month = 4,
	publisher = {{IOP} Publishing},
	volume = {15},
	number = {3},
	pages = {031005},
	author = {F Lotte and L Bougrain and A Cichocki and M Clerc and M Congedo and A Rakotomamonjy and F Yger},
	title = {A review of classification algorithms for {EEG}-based brain{\textendash}computer interfaces: a 10 year update},
	journal = {Journal of Neural Engineering},
}

@article{
    Craik_2019,
	doi = {10.1088/1741-2552/ab0ab5},
	url = {https://doi.org/10.1088%2F1741-2552%2Fab0ab5},
	year = 2019,
	month = 4,
	publisher = {{IOP} Publishing},
	volume = {16},
	number = {3},
	pages = {031001},
	author = {Alexander Craik and Yongtian He and Jose L Contreras-Vidal},
	title = {Deep learning for electroencephalogram ({EEG}) classification tasks: a review},
	journal = {Journal of Neural Engineering},
}

@article{
    Roy_2019,
	doi = {10.1088/1741-2552/ab260c},
	url = {https://doi.org/10.1088%2F1741-2552%2Fab260c},
	year = 2019,
	month = 8,
	publisher = {{IOP} Publishing},
	volume = {16},
	number = {5},
	pages = {051001},
	author = {Yannick Roy and Hubert Banville and Isabela Albuquerque and Alexandre Gramfort and Tiago H Falk and Jocelyn Faubert},
	title = {Deep learning-based electroencephalography analysis: a systematic review},
	journal = {Journal of Neural Engineering},
}

@article{
    BCI_Competition_III,
    author = {Benjamin Blankertz and Klaus-robert Müller and Gabriel Curio and Theresa M. Vaughan and Gerwin Schalk and Jonathan R. Wolpaw and Alois Schlögl and Christa Neuper and Gert Pfurtscheller and Thilo Hinterberger and Michael Schröder and Niels Birbaumer},
    title = {The BCI competition 2003: Progress and perspectives in detection and discrimination of EEG single trials},
    journal = {IEEE TRANS. BIOMED. ENG},
    year = {2004},
}

@article{
    NER_2015,
    author = {Margaux, Perrin and Emmanuel, Maby and S\'{e}bastien, Daligault and Olivier, Bertrand and J\'{e}r\'{e}mie, Mattout},
    title = {Objective and Subjective Evaluation of Online Error Correction during P300-Based Spelling},
    year = {2012},
    issue_date = {January 2012},
    publisher = {Hindawi Limited},
    address = {London, GBR},
    volume = {2012},
    issn = {1687-5893},
    url = {https://doi.org/10.1155/2012/578295},
    doi = {10.1155/2012/578295},
    journal = {Adv. in Hum.-Comp. Int.},
    month = jan,
    articleno = {Article 4},
}

@article{
    bnci-horizon-2020-8,
    AUTHOR={Riccio, Angela and Simione, Luca and Schettini, Francesca and Pizzimenti, Alessia and Inghilleri, Maurizio and Olivetti Belardinelli, Marta and Mattia, Donatella and Cincotti, Febo},
    TITLE={Attention and P300-based BCI performance in people with amyotrophic lateral sclerosis},
    JOURNAL={Frontiers in Human Neuroscience},
    VOLUME={7},
    PAGES={732},
    YEAR={2013},
    URL={https://www.frontiersin.org/article/10.3389/fnhum.2013.00732},
    DOI={10.3389/fnhum.2013.00732},
    ISSN={1662-5161},
}

@article{
    EPFL_P300_dataset,
    title = {An efficient P300-based brain-computer interface for  disabled subjects},
    author = {Hoffmann, Ulrich and Vesin, Jean-Marc and Ebrahimi,  Touradj and Diserens, Karin},
    journal = {Journal of Neuroscience Methods},
    number = {1},
    volume = {167},
    pages = {115-125},
    year = {2008},
    note = {Datasets and MATLAB-Code are available at  http://bci.epfl.ch},
    url = {http://infoscience.epfl.ch/record/101093},
    doi = {10.1016/j.jneumeth.2007.03.005},
}

@article{
    MOABB_package,
	doi = {10.1088/1741-2552/aadea0},
	url = {https://doi.org/10.1088%2F1741-2552%2Faadea0},
	year = 2018,
	month = {9},
	publisher = {{IOP} Publishing},
	volume = {15},
	number = {6},
	pages = {066011},
	author = {Vinay Jayaram and Alexandre Barachant},
	title = {{MOABB}: trustworthy algorithm benchmarking for {BCIs}},
	journal = {Journal of Neural Engineering},
}

@article{
    EDF_format,
    author = {Kemp, Bob and Olivan, Jesus},
    year = {2003},
    month = {10},
    pages = {1755-61},
    title = {European data format 'plus' (EDF+), an EDF alike standard format for the exchange of physiological data},
    volume = {114},
    journal = {Clinical neurophysiology : official journal of the International Federation of Clinical Neurophysiology},
    doi = {10.1016/S1388-2457(03)00123-8},
}

@article{
    PyRiemann,
    author    = {Marco Congedo and Alexandre Barachant and Anton Andreev},
    title     = {A New Generation of Brain-Computer Interface Based on Riemannian Geometry},
    journal   = {CoRR},
    volume    = {abs/1310.8115},
    year      = {2013},
    url       = {http://arxiv.org/abs/1310.8115},
    archivePrefix = {arXiv},
    eprint    = {1310.8115},
    bibsource = {dblp computer science bibliography, https://dblp.org},
}

@article{
    SciPy,
    author = {
        {Virtanen}, Pauli and {Gommers}, Ralf and {Oliphant},
        Travis E. and {Haberland}, Matt and {Reddy}, Tyler and
        {Cournapeau}, David and {Burovski}, Evgeni and {Peterson}, Pearu
        and {Weckesser}, Warren and {Bright}, Jonathan and {van der Walt},
        St{\'e}fan J.  and {Brett}, Matthew and {Wilson}, Joshua and
        {Jarrod Millman}, K.  and {Mayorov}, Nikolay and {Nelson}, Andrew
        R.~J. and {Jones}, Eric and {Kern}, Robert and {Larson}, Eric and
        {Carey}, CJ and {Polat}, {\.I}lhan and {Feng}, Yu and {Moore},
        Eric W. and {Vand erPlas}, Jake and {Laxalde}, Denis and
        {Perktold}, Josef and {Cimrman}, Robert and {Henriksen}, Ian and
        {Quintero}, E.~A. and {Harris}, Charles R and {Archibald}, Anne M.
        and {Ribeiro}, Ant{\^o}nio H. and {Pedregosa}, Fabian and
        {van Mulbregt}, Paul and {Contributors}, SciPy 1. 0
    },
    title = "{SciPy 1.0: Fundamental Algorithms for Scientific Computing in Python}",
    journal = {Nature Methods},
    year = "2020",
    volume={17},
    pages={261--272},
    adsurl = {https://rdcu.be/b08Wh},
    doi = {https://doi.org/10.1038/s41592-019-0686-2},
}

@inproceedings{
    Brain_Invaders_2011,
    TITLE = {{''Brain Invaders'': a prototype of an open-source P300- based video game working with the OpenViBE platform}},
    AUTHOR = {Congedo, Marco and Goyat, Matthieu and Tarrin, Nicolas and Ionescu, Gelu and Varnet, L{\'e}o and Rivet, Bertrand and Phlypo, Ronald and Jrad, Nisrine and Acquadro, Michael and Jutten, Christian},
    URL = {https://hal.archives-ouvertes.fr/hal-00641412},
    BOOKTITLE = {{5th International Brain-Computer Interface Conference 2011 (BCI 2011)}},
    ADDRESS = {Graz, Austria},
    PAGES = {280-283},
    YEAR = {2011},
    MONTH = Sep,
    HAL_ID = {hal-00641412},
    HAL_VERSION = {v1},
}

@article{
    Plug_and_Play_P300_BCI,
    author    = {Alexandre Barachant and Marco Congedo},
    title     = {A Plug{\&}Play {P300} {BCI} Using Information Geometry},
    journal   = {CoRR},
    volume    = {abs/1409.0107},
    year      = {2014},
    url       = {http://arxiv.org/abs/1409.0107},
    archivePrefix = {arXiv},
    eprint    = {1409.0107},
    bibsource = {dblp computer science bibliography, https://dblp.org}
}

@incollection{
    Recreational_Applications,
    TITLE = {{Recreational Applications of OpenViBE: Brain Invaders and Use-the-Force}},
    AUTHOR = {Andreev, Anton and Barachant, Alexandre and Lotte, Fabien and Congedo, Marco},
    URL = {https://hal.archives-ouvertes.fr/hal-01366873},
    BOOKTITLE = {{Brain-Computer Interfaces 2: Technology and Applications}},
    EDITOR = {Maureen Clerc and Laurent Bougrai and Fabien Lotte},
    PUBLISHER = {{John Wiley}},
    VOLUME = {chap. 14},
    PAGES = {241-257},
    YEAR = {2016},
    MONTH = Aug,
    HAL_ID = {hal-01366873},
    HAL_VERSION = {v1},
}

@article{
    MI_dataset,
    author = {Kaya, Murat and Binli, Mustafa and Ozbay, Erkan and Yanar, Hilmi and Mishchenko, Yuriy},
    year = {2018},
    month = {10},
    pages = {180211},
    title = {A large electroencephalographic motor imagery dataset for electroencephalographic brain computer interfaces},
    volume = {5},
    journal = {Scientific Data},
    doi = {10.1038/sdata.2018.211},
}

@article{
    How_many_people_are_able,
    author = {Guger, Christoph and Daban, Shahab and Sellers, Eric and Holzner, Clemens and Krausz, Gunther and Carabalona, Roberta and Gramatica, Furio and Edlinger, Günter},
    year = {2009},
    month = {09},
    pages = {94-8},
    title = {How many people are able to control a P300-based brain-computer interface (BCI)?},
    volume = {462},
    journal = {Neuroscience Letters},
    doi = {10.1016/j.neulet.2009.06.045},
}

@article{
    BCI_competition_III_svm,
    author = {Rakotomamonjy, Alain and Guigue, Vincent},
    year = {2008},
    month = {04},
    pages = {1147 - 1154},
    title = {BCI competition III: dataset II- ensemble of SVMs for BCI P300 speller},
    volume = {55},
    journal = {Biomedical Engineering, IEEE Transactions on},
    doi = {10.1109/TBME.2008.915728},
}

@article{
    MNE_package,
    AUTHOR={Gramfort, Alexandre and Luessi, Martin and Larson, Eric and Engemann, Denis and Strohmeier, Daniel and Brodbeck, Christian and Goj, Roman and Jas, Mainak and Brooks, Teon and Parkkonen, Lauri and Hämäläinen, Matti},
    TITLE={MEG and EEG data analysis with MNE-Python},
    JOURNAL={Frontiers in Neuroscience},
    VOLUME={7},
    PAGES={267},
    YEAR={2013},
    URL={https://www.frontiersin.org/article/10.3389/fnins.2013.00267},
    DOI={10.3389/fnins.2013.00267},
    ISSN={1662-453X},   
}

@article{
    Kaplan_Shishkin_games,
    author={A. Y. {Kaplan} and S. L. {Shishkin} and I. P. {Ganin} and I. A. {Basyul} and A. Y. {Zhigalov}},
    journal={IEEE Transactions on Computational Intelligence and AI in Games},
    title={Adapting the P300-Based Brain–Computer Interface for Gaming: A Review},
    year={2013},
    volume={5},
    number={2},
    pages={141-149},
}

\end{document}